\documentclass [12pt]{article}
\usepackage{lscape}                                         
\usepackage[centertags]{amsmath}
\usepackage{amsfonts} \usepackage{amssymb}\usepackage{eufrak}
\usepackage{graphicx}
\usepackage{a4}
\usepackage{cite}
\usepackage{lscape}

\begin{document}
\begin{titlepage}
\rightline{}
\rightline{}
\vskip 3.0cm
\centerline{\LARGE \bf  Two Double String Theory Actions:}
\vskip 0.5cm
\centerline{\LARGE \bf non-covariance vs. covariance \footnote{Talk presented at the ``2014 Corfu Summer Institute - Workshop on Quantum Fields and Strings''  (Corfu, Greece -September 14-21 2014) and based on ref. \cite{DGMP}: L. De Angelis, G. Gionti S. J., R. Marotta and F. P., {\em Comparing Double String Theory Actions}, JHEP 04 (2014) 17.} }

\vskip 1.0cm \centerline{\bf F. Pezzella}
\vskip .6cm
\vskip .4cm
\centerline{\sl Istituto Nazionale di Fisica Nucleare, Sezione di Napoli}
\centerline{ \sl Complesso Universitario di Monte
S. Angelo ed. 6, via Cintia,  80126 Napoli, Italy}
\vskip 0.4cm

\begin{abstract}
The aim of this work is to achieve a formulation of the  bosonic string theory in which T-duality appears as a manifest symmetry. Two models satisfying this requirement are discussed. The first is based on the stringy extension of the Floreanini-Jackiw Lagrangians for chiral fields and is characterized by the loss of manifest covariance on the world-sheet, recovered on mass-shell. The second model exhibits a manifest two-dimensional invariance while doubling {\em a priori} the string coordinates in the target-space. They are shown to be equivalent.

\end{abstract}

\end{titlepage}

\newpage

\section{Introduction and Motivation}
In (bosonic) string theory, the presence of compact dimensions implies the existence of two different kinds of modes: the momentum modes $p^{a}= k^{a}/R$ which are quantized ($k^{a} \in Z) $ along such dimensions and the winding modes $w^{a}$. The latter are defined  through the periodicity condition of a closed string along  a compact dimension, $ X^{a} (\tau, \sigma+ \pi) = X^{a} (\tau, \sigma) + 2 \pi R w^{a} $ with $w^{a} \in Z$  and represent the number of times the string winds around it. On the other hand, to each compact dimension $X^{a} = X^{a}_{L}(\tau + \sigma) + X^{a}_{R}(\tau - \sigma)$ one can associate the corresponding T-dual dimension  $\tilde{X}^{a} \equiv X^{a}_{L}(\tau + \sigma) - X^{a}_{R}(\tau - \sigma)$ obtained by sending:
\[
X^{a}_{L} \rightarrow X^{a}_{L}   \,\,\,\,\,\,\,\, ;  \,\,\,\,\,\,\, X^{a}_{R} \rightarrow - X^{a}_{R}  \,\, .
\]
In this way,  the winding number $w^{a}$ becomes the momentum associated with the dual coordinate.

T-duality is an old subject in string theory (for a recent review, see ref. \cite{1302.1719}): it implies that in many cases two different geometries for the extra-dimensions are physically equivalent.  It constitutes an exact symmetry for the bosonic closed string and, in the case of the simplest  compactification of one coordinate  on a circle of radius $R$ (and in the presence of only the background metric), is encoded by the following transformations:
\[
R \leftrightarrow \frac{\alpha'}{R}   \,\,\,\,\,\,\,\, ;  \,\,\,\,\,\,\,  k \leftrightarrow w 
\]
which leave the string mass spectrum invariant. The interchange of $w$ and $k$ means that the momentum excitations in one description correspond to the winding mode excitations in the dual description and viceversa.  T-duality symmetry is a clear indication that ordinary geometric concepts can break down in string theory at the string scale.  

In a background made by a tensor metric $G$ and a Kalb-Ramond field $B$, the string living on a $d$-dimensional torus exhibits a T-duality group $O(d,d;Z)$. This symmetry is reminescent of the duality $O(D,D)$ appearing at the classical level as a continuos symmetry  in  $D$ non-compact spacetime dimensions. After the compactification of  $d=D-n$ dimensions, then it breaks into $O(n,n) \otimes O(d,d;Z)$.  

The $O(D,D)$ symmetry can be explicitly inferred from the string Hamiltonian and the constraints of the theory \cite{R}. 

 Consider the usual action of a bosonic string in a $(G,B)$-background and in the conformal gauge:
 \begin{eqnarray}
S= \frac{T}{2} \int_{\Sigma} \left[ G_{\mu \nu} \, dX^{\mu} \wedge *dX^{\nu} + B_{\mu \nu} \, dX^{\mu} \wedge dX^{\nu} \right]  \label{action}
\end{eqnarray}
where the Hodge operator is defined in terms  of the metric $h= \mbox{diag} (-1,1)$ of the two-dimensional world-sheet  $\Sigma$.  The equation of motion for the coordinate $X^{\mu}$ generated by $S$ is the following:
\begin{eqnarray}
d*dX^{\mu} + \Gamma^{\mu}_{\, \nu \rho} dX^{\nu} \wedge * dX^{\rho} = \frac{1}{2} G^{\mu \sigma}H_{\sigma \nu \rho} dX^{\nu} \wedge dX^{\rho}
\end{eqnarray}
being  $ \Gamma^{\mu}_{\, \nu \rho} $ the coefficients of the Levi-Civita connection defined on the  tangent space of the $D$-dimensional target space in which the string moves.

The equation of motion for the world-sheet metric implies, as it is well-known, the vanishing of the energy-momentum tensor which turns out to give the following constraints:

\begin{eqnarray}
G_{\mu \nu} (\dot{X}^{\mu} \dot{X}^{\nu} + X'^{\mu} X'^{\nu} ) = 0 \,\,\,\,\,\,\,\, ; \,\,\,\,\,\,\,\,  G_{\mu \nu} \dot{X}^{\mu} X'^{\nu}=0    \label{constr}
\end{eqnarray}
with the usual convention of defining $\dot{X} \equiv \partial_{\tau} X$ and $X' \equiv \partial_{\sigma}X$. After introducing the canonical momentum $P_{\mu}$ conjugate to the coordinate $X^{\mu}$:
\begin{eqnarray}
P_{\mu} = \frac{\partial L}{\partial \dot{X}^{\mu}} = T \left( G_{\mu \nu} \, \dot{X}^{\nu} + B_{\mu \nu} X'^{\nu} \right) ,
\end{eqnarray}
once can rewrite the Hamiltonian in terms of a generalized 2D-dim vector 
\begin{eqnarray}
A^{i} = \left( \begin{array}{c}
\partial_{\sigma} X^{\mu} \\  
2 \pi \alpha' P_{\mu} 
\end{array} \right)  \,\,\,\,\,\,\,\,\,\,\,\,\,\,\,\,\, i=1, \cdots, 2D
 \end{eqnarray}
  as follows:
\begin{eqnarray}
H(X; G, B) = \frac{1}{4 \pi \alpha'} \, A^{t} \, M \, A
\end{eqnarray}
where $M$ is the so-called {\em generalized metric} \cite{D,T}:
\begin{eqnarray}
M= \left( \begin{array}{cc}
G-BG^{-1}B & BG^{-1} \\
-G^{-1}B & G^{-1} 
\end{array} \right) .
\end{eqnarray}

The constraints can also be expressed in terms of $M$. In particular, the second constraint in Eq. (\ref{constr}) can be rewritten as $A^{t} M A=0$ and sets the Hamiltonian to zero, while the dynamic is completely governed by the first constraint in Eq. (\ref{constr}) rewritten as:
\begin{eqnarray}
A^{t} \Omega A=0  \,\,\,\,\,\,\,\,\,\,  \mbox{with} \,\,\,\,\,\,\,\,\,\,\, \Omega= \left( \begin{array}{cc}
0 & 1 \\
1 & 0 
\end{array} \right) .
\end{eqnarray}
Here $\Omega$ is the metric left invariant by an element of the group $O(D,D)$  defined by the matrices  $T$ satisfying the condition $T^{t} \Omega T = \Omega$. Hence, all of the  generalized vectors solving the constraints are related by $O(D,D)$ transformations. This shows how the $O(D,D)$ invariance emerges very naturally in the bosonic string model.

In particular, in the presence of constant $G$ and $B$ as background, the equations of motion for the string coordinates provide a set of conservation laws on the world-sheet, written in terms of conserved corrents $J_{\mu}^{\alpha}$ here defined:

\begin{eqnarray}
J^{\alpha}_{\mu}  = G_{\mu \nu} \partial^{\alpha} X^{\nu} + \epsilon^{\alpha \beta} B_{\mu \nu} \partial_{\beta} X^{\nu} \,\,\,\,\,\,\,\,\,\,\,\,\,\, \mbox{with} \,\,\,\,\,\,\,\,\,\,\,
\partial_{\alpha} \, J^{\alpha}_{\mu}=0   \end{eqnarray}
which, locally,  can be expressed in terms of a new coordinate $\tilde{X}$ through the identification:
\begin{eqnarray}
J^{\alpha}_{\mu} \equiv - \epsilon^{\alpha \beta} \partial_{\beta} \tilde{X}_{\mu} .  \label{tildex}
\end{eqnarray}
Eq. (\ref{tildex}) provides the definition of the dual coordinate $\tilde{X}_{\mu}$ of $X^{\mu}$ and 
 reproduces, for $B=0$, the known duality relation between $X$ and $\tilde{X}$:

\[ G_{\mu \nu} \partial^{\alpha} X^{\nu} \equiv - \epsilon^{\alpha \beta} \partial_{\beta} \tilde{X}_{\mu} \,\, .
\]
In terms of the field $\tilde{X}$, the dual action $\tilde{S}$  reads as:
\begin{eqnarray}
\tilde{S}= \frac{T}{2} \int_{\Sigma} \left[ \tilde{G}_{\mu \nu} \, d\tilde{X}^{\mu} \wedge *d \tilde{X}^{\nu} + \tilde{B}_{\mu \nu} \, d\tilde{X}^{\mu} \wedge d\tilde{X}^{\nu} \right]   \,\, 
\end{eqnarray}
and the two actions, $S$ and $\tilde{S}$  are obtained from each other under the exchanges:
\begin{eqnarray}
G \leftrightarrow \tilde{G} = G-BG^{-1}B  \,\,\,\,\,\,\,\,\,\,\,\,\,\,\,\,\,\,  \,\,\,\,\,\,\,\,\,\,\,\,\,\,\,  B \leftrightarrow \tilde{B}=- \tilde{G} B^{-1} G
\end{eqnarray}
corresponding to the following transformation of the generalized metric:
\begin{eqnarray}
M \rightarrow M^{-1}  \,\, .
\end{eqnarray}

It is interesting to note that the equations of motion for the coordinates $(X, \tilde{X})$ can be combined into the following single equation invariant under $O(D,D)$ transformations: 
\begin{eqnarray}
M \partial_{\alpha} \chi =  \Omega\,\,  \epsilon_{\alpha \beta} \,\, \partial^{\beta} \chi \,\,\,\,\,\,\,\,\,\,\, \mbox{with} \,\,\,\,\,\,\,\,\,\,\,\,\,\,\,  \chi \equiv \left( \begin{array}{c}  X \\ \tilde{X} \end{array} \right) \,\,\, .
\end{eqnarray}

If $d$ space coordinates are compactified on a $d$-torus $T^{d}$, then the dual coordinate $\tilde{X}$ will satisfy the same periodicity conditions as $X$.  For closed strings such compactification means the following periodicity conditions:
\[
X^{a}(\sigma, \tau) = X^{a} (\sigma + 2 \pi , \tau) + 2 \pi L^{a} \,\,\,\,\,\,\,\,   {\mbox with}  \,\,\,\,\,\, L^{a}= \sum_{i=1}^{d} w_{i} R_{i} e^{a}_{i}
\]
where $w_{i}$ is  the winding number and the vectors $e^{a}_{i}$ with $i=1, \cdots, d$ form a basis on $T^{d}$. 
Therefore,  the continuous $O(D,D)$ duality generates the  duality {\em symmetry} $O(d,d;Z)$ on the torus.

These features make it tempting to construct a theory having an $O(D,D)$ invariance at the very beginning. 

A sigma model satisfying this requirement was proposed by A. A. Tseytlin \cite{T} and in it
the $O(D,D)$ invariance is manifested by introducing the set of coordinates $(X^{\mu}, \tilde{X}_{\mu})$ (with $\mu=1, \dots, D$) in the {\em doubled} $2D$-dimensional target space \cite{D}.  
Recent developments of Double Field Theory \cite{S, DFT, HLZ, DFT2} have the same inspirations. The world-sheet sigma model proposed by Tseytlin does not have manifest covariance but this is recovered on-shell.

 Another model, due to C. Hull  \cite{H}, exhibits a manifest two-dimensional invariance while doubling {\em a priori} the string coordinates in the target-space. More precisely, Hull starts with a covariant action involving  a double number of coordinates and the $O(D,D)$ invariance is generated when a self-duality constraint, halving the degrees of freedom, is imposed.  
 
 The classical and quantum properties of these two models  are going to be discussed in this work, showing their equivalence.

Further aspects of double string theory actions have been studied by other authors \cite{BT, NP,  C, LP}.

As a consequence of doubling the space-time coordinates, the spectrum states of a closed string must be described by vertex operators which must include 
 $(X, \tilde{X})$. This occurs also for the winding modes which appear in the spectrum of a closed string living on a torus. The torus vacuum can therefore be perturbated by such operators,  leading to other possible vacua corresponding to two-dimensional field theories with interactions depending on $(X, \tilde{X})$.  In the underlying two-dimensional local Quantum Field Theory one must treat $(X, \tilde{X})$ as independent scalar fields which appear to be dual to each other on the mass-shell and in the absence of interactions.
 
 The first step of this program would be  to find the Lagrangian of the free scalar two-dimensional fields $X$ and $\tilde{X}$ and this must describe $D$ and not $2D$ degrees of freedom in order to establish a correspondence with the standard formulation. The second step would be to introduce the possible interaction terms and to determine, 
 through the Renormalization Group flow equations, which of them, in particular,  correspond to string vacua. Furthermore, in the extended formulation,  the symmetry under $X \leftrightarrow \tilde{X}$ must be implemented as an off-shell symmetry of the world-sheet action. This makes the duality invariance of the scattering amplitudes and the effective action 
manifest.

From the extended formulation one can recover the known results for the partition function and the scattering amplitudes of the usual formulation - which involves only the coordinates $X$ - under the hypothesis that the interaction terms would not depend on $\tilde{X}$. In this way once can integrate $\tilde{X}$ out in the path-integral of the double theory, reproducing  the usual results for the partition function and the scattering amplitudes. This happens when the compactification radii are much larger than the string length: in this case, indeed, the winding modes result to be massive so that the relevant interactions are only dependent on $X$. At intermediate scales at which the compactification radii are of the same scale as the string length, the interaction terms are dependent on both the coordinates $X$ and $\tilde{X}$. At small scales, when compactification radii are much smaller than the string length, the relevant interactions are dependent on $\tilde{X}$.

The above considerations show that the {\em T-duality symmetric formulation of string theory can be considered a natural generalization of the standard formulation at the string scale}.

\section{A Simple Model: Free Scalar Field Theory in 2 dimensions}

Before discussing the double string model proposed by Tseytlin \cite{T}, it is interesting to consider first the usual Lagrangian of a two-dimensional scalar field $\phi$  with the aim of rewriting it in a manifestly invariant form under the exchange of $\phi$ with its Hodge dual $\tilde{\phi}$. The starting point is therefore: 

\begin{eqnarray}
{\cal L} = -\frac{1}{2} \partial_{a} \phi  \partial^{a} \phi = \frac{1}{2} \left( \dot{\phi}^{2} - \phi'^{2} \right) \,\,\,\,\,\,\,\,\,\, \mbox{with} \,\,\,\,\,\,\,  \eta_{ab}= \left( \begin{array}{cc} -1 & 0 \\ 0 & 1 \end{array} \right) \,\, \label{scalag}
\end{eqnarray}
giving the free two-dimensional wave equation as equation of motion for $\phi$:
\begin{eqnarray}
( \partial^{2}_{\sigma} - \partial^{2}_{\tau} ) \phi = 0  .   \label{eqphi}
\end{eqnarray}
It is well-known that such a field admits a Hodge dual defined by:
\begin{equation}
\partial_{a} \phi = - \epsilon_{ab} \partial^{b} \tilde{\phi} \,\,\,\,\,\,\,\,\,\,\,\, \mbox{or, equivalently} \,\,\,\,\,\,\,\,\, *d\phi= d\tilde{\phi} \, .   \label{duacon}
\end{equation}
and satisfying the same equation of motion as $\phi$ in (\ref{eqphi}). Therefore, the exchange $\phi \leftrightarrow \tilde{\phi}$ is a symmetry of the theory.

The  Lagrangian density ${\cal L}$  can be rewritten in such a way that the two fields $\phi$ and $\tilde{\phi}$ appear on equal footing with a manifest invariance under $\phi \leftrightarrow \tilde{\phi}$. 
 In order to achieve this goal, two steps have to be made. The first consists in rewriting ${\cal L}$ in a first order form as follows, after introducing the auxiliary field $p$:
 \begin{eqnarray}
 {\cal L} \rightarrow {\cal L}' [p,\phi] = p \dot{\phi} - \frac{1}{2} p^{2} - \frac{1}{2} \phi'^{2}  \,\,\,\,\,\,\, \mbox{providing} \,\,\,\,\,\,   p=\dot{\phi} \,\,. 
  \end{eqnarray}
  The second step consists of introducing a Lagrange multiplier $b$ that takes into account the constraint $\dot{\phi} = \phi'$:
\begin{eqnarray}
{\cal L'} \rightarrow {\cal L''} = p \dot{\phi} - \frac{1}{2} p^{2} - \frac{1}{2} \phi'^{2} +b (p- \phi')  . \nonumber  
\end{eqnarray}
Solving the constraint provides the Lagrangian density:
\begin{eqnarray}
{\cal L''} [\phi, \tilde{\phi}] = \dot{\phi} \tilde{\phi}' - \frac{1}{2} \dot{\phi}^{2} - \frac{1}{2} \phi'^{2}  \nonumber 
\end{eqnarray}
that in turn, after an integration by parts, can be rewritten in the following form symmetric under $\phi \leftrightarrow \tilde{\phi}$:  
\begin{eqnarray}
{\cal L}_{sym} = \frac{1}{2} \dot{\phi} \tilde{\phi}' + \frac{1}{2} \phi' \dot {\tilde{\phi}} - \frac{1}{2} \phi'^{2} -\frac{1}{2} \tilde{\phi'}^{2} \,\, .
\end{eqnarray}
 The first observation is that the requirement of the latter symmetry has destroyed the Lorentz invariance, but it will be demonstrated soon that this is recovered on-shell. 
The equations of motion for $\phi$ and ${\tilde{\phi}}$  derived from ${\cal L}_{sym}$ are:
\begin{eqnarray}
\phi \rightarrow  \partial_{\sigma} \left[ \partial_{\sigma} \phi - \partial_{\tau} \tilde{\phi} \right] = 0 \nonumber  \\
\tilde{\phi} \rightarrow \partial_{\sigma} \left[ \partial_{\sigma} \tilde{\phi} - \partial_{\tau} \phi \right] = 0 \nonumber
\end{eqnarray}
satisfied, respectively, by the solutions:
\begin{eqnarray}
\partial_{\sigma} \phi - \partial_{\tau} \tilde{\phi} = f( \tau ) \nonumber \\
\partial_{\sigma} \tilde{\phi} - \partial_{\tau} \phi = \tilde{f} ( \tau)   \label{sol}
\end{eqnarray}
At this point one can invoke another symmetry of ${\cal L}_{sym}$ under the transformations:
\begin{eqnarray}
\phi \rightarrow \phi + g(\tau)   \,\,\,\,\,\,\,\, ; \,\,\,\,\,\, \tilde{\phi}  \rightarrow  \tilde{\phi} + \tilde{g} (\tau)
\end{eqnarray}
so that the two functions $f(\tau)$ and $f(\tilde{\tau})$ can be gauged to zero and in this way the solutions reproduce {\em on-shell} the duality conditions (\ref{duacon}).

Furthermore, one can observe that also the Lorentz invariance is recoverd on-shell. In fact, ${\cal L}_{sym}$ is invariant under the transformation:
\begin{eqnarray}
\delta \phi = \tau \phi' + \sigma \tilde{\phi}'  \,\,\,\,\,\,\,\, \mbox{and} \,\,\,\,\,\,\,\,\,\,\,
\delta \tilde{\phi} = \tau \tilde{\phi}' + \sigma \dot{\phi}
\end{eqnarray}
which, if  the duality conditions $\dot{\phi}=\tilde{\phi}'$ and $\phi'= \dot{\tilde{\phi}}$ (valid on-shell, as above shown) are taken into consideration, reproduce  the usual two-dimensional Lorentz rotations on-shell.

The symmetric Lagrangian ${\cal L}_{sym}$ can be diagonalized by introducing a pair of scalar fields $\phi_{+}$ and $\phi_{-}$ defined by:
\begin{eqnarray}
\phi \equiv \frac{1}{\sqrt{2}} \left( \phi_{+} + \phi_{-} \right) \,\,\,\,\,\,\,\, ; \,\,\,\,\,\,\,\, \tilde{\phi} \equiv \frac{1}{\sqrt{2}} \left( \phi_{+} - \phi_{-} \right) 
\end{eqnarray}
in terms of which it becomes the sum of two Lagrangians, one depending on $\phi_{+}$ and the other on $\phi_{-}$:
\begin{equation}
{\cal L}_{sym} = {\cal L}_{+}(\phi_{+}) + {\cal L}_{-}(\phi_{-})   \label{FJ}
\end{equation}
with 
\begin{eqnarray}
{\cal L}_{\pm} (\phi_{\pm}) = \pm \frac{1}{2} \dot{\phi}_{\pm} \phi'_{\pm} - \frac{1}{2} \phi'^{2}_{\pm} \,\, .  \label{FJL}
\end{eqnarray}
Here,  ${\cal L}_{+}$ and ${\cal L}_{-}$ are the Floreanini-Jackiw  Lagrangian densities \cite{FJpap} for chiral and antichiral fields. In fact, the chirality and the chirality conditions, respectively for $\phi_{+}$ and $\phi_{-}$, are obtained by solving the equations of motion that one can derive from (\ref{FJ}):
\begin{eqnarray}
\phi_{+} \rightarrow   \partial_{\sigma} \left[ \dot{\phi} - \phi'_{+} \right]=0  \,\,\,\,\,\,\,\, ; \,\,\,\,\,\, \phi_{-} \rightarrow \partial_{\sigma} \left[ \dot{\phi}_{+} + \phi'_{-} \right] = 0  \label{eqofmot}
\end{eqnarray}
and noticing that, as already seen for the Lagrangian density ${\cal L}_{sym}$, due to the symmetry under
\begin{equation}
\phi_{+} \rightarrow \phi_{+} + g(\tau)  \,\,\,\,\, ; \,\,\,\,\,  \phi_{-} \rightarrow \phi_{-} + \tilde{g} (\tau) \,\,\, ,  \label{shift}
\end{equation}
one can gauge to zero the arbitrary functions $f(\tau)$ and $\tilde{f}(\tau)$ appearing in the solutions of the equations of motion (\ref{eqofmot}):
\[
\dot{\phi}_{+} - \phi'_{+} = f(\tau)  \,\,\,\,\,\,\,\,  ; \,\,\,\,\,\,\,  \dot{\phi} + \phi'_{-} = \tilde{f} (\tau) \,\,\, .
\]

The Floreanini-Jackiw Lagrangians are first-order and, in the case of a discrete number of degrees of freedom $q^{i}$ with $i=1, \cdots, N$, they look like:
\begin{eqnarray}
 L = \frac{1}{2} q^{i} c_{ij} \dot{q}^{j} - V(q)  \,\,\,\,\,\, \mbox{with} \,\,\,\,\, \mbox{det} \,\, c_{ij} \neq 0 \,\, .  \label{FJ2}
\end{eqnarray}
They are characterized by $N$ primary constraints:
\[
T_{j} \equiv p_{j} - \frac{1}{2} q^{i} c_{ij} \,\, ,
\]
where $p_{j}$ is the momentum canonically conjugate to $q^{j}$. These are second class constraints, since $Q_{ij} \equiv \left\{ T_{i}, T_{j} \right\}_{PB} = c_{ij} \neq 0$ .

In order to quantize the theory described classically by the Lagrangian (\ref{FJ2}), the Dirac quantization method has to be applied with the corresponding brackets. It is worth reminding that the Dirac bracket of any two functions of the phase space variables is defined in terms of the corresponding Poisson brackets as follows:
\[
\left\{ f,g \right\}_{DB} \equiv \left\{ f,g \right\}_{PB} - \left\{ f, T_{j} \right\}_{PB} (Q^{-1})_{jk} \left\{ T_{k},g \right\}_{PB} \,\,\, .
\]
According to the usual transition rule $i \left\{ f,g \right\}_{DB} \rightarrow \left\{ f,g \right\} $ from the classical to the quantum theory,  the following commutators are obtained:
\begin{eqnarray}
\left[ q_{i}, q_{j} \right] = i c^{-1}_{ij}  \,\,\,\,\,\,;\,\,\,\,\,\,  \left[ q_{i}, p_{j} \right] = \frac{1}{2} i \,\,\delta_{ij} \,\,\,\,;\,\,\,\,\,\, \left[ p_{i}, p_{j} \right] = - \frac{1}{4} i c_{ij} \,\, .
\end{eqnarray}
The two Lagrangian densities ${\cal L}_{+}$ and ${\cal L}_{-}$ can be rewritten in such a way to put in evidence the structure in eq. (\ref{FJ2}) by making, respectively, the following correspondences:
\[ 
q^{i} \rightarrow \chi(\sigma) = \pm \phi'_{\pm}(\sigma)
\]
together with 
\[
V(q) \rightarrow \frac{1}{2} \int d \sigma \chi^{2}(\sigma) \,\,\,\,\,  ; \,\,\,\,\, c_{ij} \rightarrow \frac{1}{2} \epsilon (\sigma- \sigma') \,\,\,\,\,;\,\,\,\,\, \frac{1}{2} \partial_{\sigma} \epsilon (\sigma - \sigma') = \delta (\sigma - \sigma') \,\,.
\]
Both of them can be put in the same following form:
\begin{eqnarray}
{\cal L}_{\pm} = \frac{1}{4} \int d \sigma d\sigma'  \chi(\sigma) \epsilon (\sigma - \sigma') \dot{\chi} (\sigma') - \frac{1}{2} \int d\sigma \chi^{2} (\sigma)
\end{eqnarray}
with the momentum conjugate to $\chi(\sigma)$ being given by
\[
\pi (\sigma, \tau) = \frac{1}{4} \int d \sigma' \chi(\sigma', \tau) \epsilon (\sigma-\sigma')
\]
while the primary second class constraints read as:
\[
\phi(\sigma, \tau) \equiv \pi(\sigma, \tau) - \frac{1}{4} \int d \sigma' \chi(\sigma', \tau) \epsilon (\sigma-\sigma') \,\, .
\]
Then, the Dirac quantization method yields straightforwardly to the following commutators:
\begin{eqnarray}
\left[ \phi_{\pm} (\sigma, \tau ), \phi_{\pm} (\sigma', \tau') \right ] & =  &\pm i \frac{1}{2} \epsilon(\sigma - \sigma')  \nonumber \\
\left[ \phi_{\pm} (\sigma, \tau ), \pi_{\pm} (\sigma', \tau') \right ]  & = &  \frac{i}{2} \delta' (\sigma- \sigma') \nonumber\\
\left[ \pi_{\pm} (\sigma, \tau ), \pi_{\pm} (\sigma', \tau') \right ]  & =  & - \frac{i}{8} \epsilon(\sigma - \sigma')  \,\, . \label{comm}
\end{eqnarray}

It is very interesting to see from the first of the above commutators that $\phi_{\pm}$ or, equivalently, $\phi, \tilde{\phi}$ behave like non commuting phase space-time coordinates. This aspect of non-commutativity emerges  very naturally in this context as a consequence of requiring the explicit symmetry of the Lagrangian density ({\ref{scalag}) under $\phi \leftrightarrow \tilde{\phi}$.

\section{Non-covariant action (Tseytlin): classical and quantum aspects.}

In this section  classical and quantum aspects of the action proposed by Tseytlin \cite{T} are discussed. It  has the property of showing an explicit $O(D,D)$ invariance but  is characterized by the loss of the Lorentz covariance that is, however, recovered on-shell.

The starting point is the Lagrangian density (\ref{FJ}) written for  $D$ scalar fields $\chi^{i}$ and in the most general background:
\begin{eqnarray}
S[ e^{a}_{\alpha}, \chi^{i}]= - \frac{1}{2} \int d^{2} \xi \, e \left[ C_{i j} \nabla_{0} \chi^{i} \, \nabla_{1} \chi^{j} + M_{i j} \nabla_{1} \chi^{i} \nabla_{1} \chi^{j} \right]   \label{act1}
\end{eqnarray}
with $i=1, \dots, D$ and where $C_{ij}$ and $M_{ij}$ are symmetric matrices and $e^{a}_{\,\, \alpha}$ ($a$ is a flat index while $\alpha$ is curved) is the zweibein appearing in the definition of the derivative covariant  of the scalar field $\chi^{i}$: $\nabla_{\alpha}  \chi^{i} \equiv e^{a}_{\,\,\alpha} \partial_{a} \chi^{i}$. 
The action (\ref{act1}) exhibits the following local invariances:
\begin{itemize}
\item{ invariance under two-dimensional diffeomorphisms $\xi^{\alpha} \rightarrow \xi'^{\alpha}( \xi)$\,\,;}
\item{
invariance under Weyl transformations $
e^{a}_{\,\, \alpha} \rightarrow \lambda (\xi) e^{a}_{\,\, \alpha}  \,\,
$} \,\, .
\end{itemize}
Furthermore, the action (\ref{act1}) must be  invariant also under the following finite transformation involving the zweibein,  since physical observables are independent on the latter:
\begin{eqnarray}
e'^{a}_{\,\,\, \alpha} = \Lambda^{a}_{\,\,b} (\xi) e^{b}_{\,\, \alpha} \,\, .
\end{eqnarray}
Here,  $\Lambda^{a}_{\,b} (\xi) $ is an arbitrary $\xi$-dependent Lorentz $SO(1,1)$ matrix. This finite transformation on $e^{a}_{\, \alpha}$  induces the following infinitesimal one:
\begin{eqnarray}
\delta e^a_{~\alpha}=  \omega^{a}_{~b} (\xi)  e^{b}_{\, \alpha}
\end{eqnarray}
with $\omega_{ab} = - \omega_{ba}$. In particular, the choice $\omega^a_{\,\, b}(\xi) =\alpha(\xi)\epsilon^{a}_{~b}$  can be performed.
But the action (\ref{act1}) is not manifestly invariant under such transformations, so the requirement of {\em on-shell} local Lorentz invariance has to be imposed and this implies the condition:
\begin{eqnarray}
\epsilon^{ab} t_{a b}=0 \,\,\,\,\, \mbox{with} \,\,\,\, t_{a}^{~b} \equiv  \frac{2}{e}  \frac{\delta S}{\delta e_{~\alpha}^{ a}} e_{~\alpha}^{b} .  \label{trzero}
\end{eqnarray}
The explicit expression for $t_{a}^{\,\,b}$ can be straightforwardly computed from the action (\ref{act1}) and it results to be:
\begin{eqnarray}
t_{a}^{\,\,b} & = & - \delta^{b}_{a} \left[ C_{ij} \nabla_{0} \chi^{i} \nabla_{1} \chi^{j} + M_{ij} \nabla_{1} \chi^{i} \nabla_{1} \chi^{j} \right] \nonumber \\
&&  +\delta_{0}^{b} C_{ij} \nabla_{a} \chi^{i} \nabla_{1} \chi^{j} + \delta_{1}^{b} C_{ij} \nabla_{0} \chi^{i} \nabla_{a} \chi^{j} + 2 \delta_{1}^{b} M_{ij} \nabla_{a} \chi^{i} \nabla_{1} \chi^{j} . \label{tab}
\end{eqnarray}
Moreover, the Weyl invariance makes the trace of tensor $t_{a}^{\,\,b} $ vanishing, i.e. $t_a^{~a} = 0$, while the equations of motion  $\frac{\delta S}{\delta e_{~\alpha}^{a}}=0$ for $e^{a}_{\alpha}$ imply $t^{~b}_{a}=0 $. 

 The invariances under diffeomorphisms and Weyl transformations, together with the local Lorentz  invariance that holds on-shell, allow to choose the {\em flat gauge} $e^{a}_{~\alpha} = \delta^{a}_{\alpha}$ for the zweibein.

In the case in which the matrices $C$ and $M$ are {\em constant}, the equation of motion for $\chi^{i}$ can be straightforwardly derived:
\begin{eqnarray}
\partial_1 \left[  C_{ij} \partial_{0} \chi^{j} + M_{ij} \partial_{1} \chi^{j} \right]  = 0 \label{1}
\end{eqnarray}
with the following surface integral:
\begin{eqnarray}
-\left. \int_{-\infty}^{+\infty} d\tau  \delta \chi^i \left( C_{ij}\partial_0\chi^j+M_{ij}\partial_1 \chi^j \right)\right|^{\sigma=\pi}_{\sigma=0} +\frac{1}{2} \left.  \int_{- \infty}^{+ \infty} d \tau \,  C_{ij} \,  \left[ \partial_{0} \chi^{j} \delta \chi^{i} \right] \right|^{\sigma=\pi}_{\sigma=0} \,\, .
\end{eqnarray}
From eq. (\ref{1}) one obtains:
\begin{eqnarray}
C_{ij} \partial_{0} \chi^{j} + M_{ij} \partial_{1} \chi^{j} = g_{i} (\tau) \label{fi}
\end{eqnarray}
being $g_{i} (\tau) $ an arbitrary $\tau$-dependent function. It is crucial, at this point, to observe that,
with  $C$ and $M$ constant,  the action (\ref{act1}) has a further local gauge symmetry under the following transformations:
\begin{eqnarray}
\chi^{i} \rightarrow \chi'^{i}= \chi^{i} + f^{i} (\tau, \sigma) \label{fursym}
\end{eqnarray}
with the functions $f^{i}$ satisfying  $ \nabla_{1} f^{i} = 0$ and the same boundary conditions as the fields $\chi$ and $\chi'$. This {\em shift symmetry}, that generalizes the one already seen in the case of a single scalar field in (\ref{shift}), leaves the equation of motion  invariant and can be used, in the flat gauge, to fix $C\,\partial_0f=g$.
As a result  one has:
\begin{eqnarray}
C_{ij} \partial_{0} \chi^{j} + M_{ij} \partial_{1} \chi^{j} = 0 \,\,  \label{fi0}
\end{eqnarray}
and the boundary conditions reduce to:
\begin{eqnarray}
\frac{1}{2} \left.  \int_{- \infty}^{+ \infty} d \tau \,  C_{ij} \,  \left[ \partial_{0} \chi^{j} \delta \chi^{i} \right] \right|^{\sigma=\pi}_{\sigma=0}=0 \,\, .
\end{eqnarray}
This term is  vanishing when  $\delta \chi^{i} =0$ (ensured by a periodicity condition in $\sigma$ imposed  on $\chi^{i}$) or, alternatively,  when $ \partial_{0} \chi^{i}=0$ at $\sigma=0, \pi$.

In the flat gauge and along the solution of the equations of motion, one can see, after some calculation, that the condition $\epsilon^{ab} t_{a b}=0$, previously discussed, implies the following relation between the matrices $C_{ij}$ and $M_{ij}$:
\begin{eqnarray}
C=MCM \,\,\,  . \label{condition}
\end{eqnarray}
If one notices that the matrix $C$ can always be  put, after suitably rotating and rescaling $\chi^{i}$, in the following diagonal form:
\begin{eqnarray}
C= \mbox{diag} (1, \cdots, 1, -1, \cdots, -1)\label{24}
\end{eqnarray}
with $p$ eigenvalues $ 1$ and $q$ eigenvalues $ -1$, then eq. (\ref{condition}) defines the indefinite orthogonal group $O(p,q)$ of $N \times N$ matrices $M$  with $N=p+q$ (with $p, q$ still undetermined at this level) in $R^{p,q}$. With this expression of $C$, it is simple to see that the action (\ref{act1}) describes two-dimensional scalar $p$  chiral and $q$ antichiral fields and the requirement of absence of a local quantum Lorentz anomaly implies $p=q=D$ with $2D=N$. Correspondingly, the action $S$ in eq. (\ref{act1}) describes a mixture of $D$ chiral scalars $\chi_{-}^{\mu}$ and $D$ antichiral scalars $\chi_{+}^{\mu}$.

It is very interesting to observe how the requirement of local Lorentz invariance imposed through the condition (\ref{trzero}) has led, in a very natural way, to the interpretation of $C$ as a metric in a 2D-dimensional target space with coordinates:
\[
\chi^{i} = (\chi_{-}^{\mu} , \chi_{+}^{\mu})  \,\,\,\,\,\,\,\,\,\,\,\, \,\,\,\,\,\,\,\,\,\,\,\,\,\, ds^{2} = d\chi^{i} \,\,\, C_{ij} \,\,\, d \chi^{j} \,\,\,\,\,\,\,\,\,\, i=1, \cdots , 2D \,\,\,\,\,  \mbox{and}\,\,\,\,\, \mu=1, \cdots , D  \,\,\, .
\]

At this point, one can make a change of coordinates in the target space by defining a set of new {\em chiral} coordinates:
\[
X^{\mu} = \frac{1}{\sqrt{2}} \left( \chi^{\mu}_{+} + \chi^{\mu}_{-} \right)  \,\,\,\,\,\,\,\,\,\,\,\,\, \tilde{X}_{\mu} = \frac{1}{\sqrt{2}} \delta_{\mu \nu} \left( \chi^{\nu}_{+} - \chi^{\nu}_{-} \right) \,\, .
\]
In these new coordinates, the $O(D,D)$ invariant metric $C$ becomes the following matrix:
\begin{eqnarray}
C_{ij} = - \Omega_{ij} = - \left( \begin{array}{cc}  \,\,\,\,\, 0_{\mu \nu}  &   \mathbb{I}_{\mu}^{\,\,\, \nu} \\ \,\,\,\,\, \mathbb{I}^{\mu}_{\,\, \nu} & \,\,\,\,\, 0^{\mu \nu} \end{array} \right)  \label{omega}
\end{eqnarray}
while the  condition (\ref{condition}) transforms into the constraint $M^{-1}~=\Omega^{-1} M\Omega^{-1}$
 on the symmetric matrix $M$ that has $D^{2}=D(D+1)/2+D(D-1)/2 $  independent elements and, therefore, it can be  parametrized by a symmetric matrix $G$ and  an antisymmetric matrix $B$.  The  expression for $M$, defined up to a sign, being the above constraint quadratic in it, is:

\begin{eqnarray} \label{M_ij}
 M_{ij}= \pm\left( \begin{array}{cc}
                                       (G-B\,G^{-1}B)_{\mu \nu} & (B\, G^{-1})_{\mu}^{\,\, \nu}\\
                                       (-G^{-1}\, B)^{\mu}_{\,\, \nu}  & (G^{-1})^{\mu \nu} \end{array}\right)  \,\, . \label{35}
\end{eqnarray}
In ref. \cite{DGMP} it has ben shown that only the positive sign of $M$  determines a positive definite Hamiltonian. Therefore, $M$   is considered positive
  in eq. (\ref{35}).
In the new system of coordinates the action (\ref{act1}) reads as:
\begin{eqnarray}
  S=\frac{1}{2} \int d^{2} \xi \,\, e \left[ \Omega_{ij} \nabla_{0} {\cal \chi}^{i} \nabla_{1} {\cal \chi}^{j} - M_{ij}  \nabla_{1} {\cal \chi}^{i} \nabla_{1} {\cal \chi}^{j} \right]. \label{actics}
  \end{eqnarray}
  It is invariant under the $O(D,D)$ transformations:
\begin{eqnarray}
{\cal \chi}'={\cal R} {\cal \chi}~~;~~M'={\cal R}^{-t}M {\cal R}^{-1}~~;~~{\cal R}^{t}\Omega {\cal R} =\Omega~~~~~~~~~~{\cal R} \in O(D,D) \,\, \label{40}
\end{eqnarray}
showing that the background itself suitably transforms.
The matrix $\Omega$  is immediately seen to belong to  $O(D,D)$ and, in particular, when ${\cal R}^i_{~j}=\Omega_{ij}$, the action (\ref{actics}), expressed in terms of $X^{\mu}$ and $\tilde{X}_{\mu}$
 \begin{eqnarray}
  S= \frac{1}{2} \int d^{2} \xi e \left[ \nabla_{0} X^{\mu} \nabla_{1} \tilde{X}_{\mu} + \nabla_{0} \tilde{X}^{\mu} \nabla_{1} X_{\mu} - (G-B\,G^{-1}B)_{\mu \nu} \nabla_{1} X^{\mu} \nabla_{1} X^{\nu} \right. \nonumber \\
 \,\,\,\,\,\,\,\,\, \left.  - \,\, (B\, G^{-1})_{\mu}^{\,\, \nu} \nabla_{1} X^{\mu} \nabla_{1} \tilde{X}_{\nu} +   (G^{-1}\, B)^{\mu}_{\,\, \nu} \nabla_{1} \tilde{X}_{\mu} \nabla_{1} X^{\nu} - (G^{-1})^{\mu \nu} \nabla_{1} \tilde{X}_{\mu} \nabla_{1} \tilde{X}_{\nu} \right]  
  \end{eqnarray}
 exhibits what in string theory will become the more familiar T-duality invariance under $X \leftrightarrow \tilde{X}$ with a consequent transformation of the generalized metric given by
 $M \rightarrow M^{-1}$. 

One can conclude that the sigma-model action 
\begin{eqnarray}
S[ e^{a}_{\alpha}, \chi^{i}]= - \frac{T}{2} \int d^{2} \xi \, e \left[ C_{i j} \nabla_{0} \chi^{i} \, \nabla_{1} \chi^{j} + M_{i j} \nabla_{1} \chi^{i} \nabla_{1} \chi^{j} \right]  \label{act1str}
\end{eqnarray}
even if non-covariant,  is the candidate to describe a bosonic string with a manifest $O(D,D)$ duality invariance. In eq. (\ref{act1str})  the usual string tension $T=\frac{1}{2 \pi l^{2}}$, where $l$ is the fundamental length of the theory, has been introduced.
In particular, if the string lives on a torus $T^{d}$,  then such symmetry becomes $O(d,d;Z)$ for the compactified dimensions and the functions $\chi^{i}$ provide the string coordinates satisfying the periodicity conditions:
\begin{eqnarray}
X^{a}(\tau,\sigma + \pi) = X^{a}(\tau, \sigma) +2\,\pi\, l\,  w^{a}~~;~~\tilde{X}_{a}(\tau, \sigma + \pi) = \tilde{X}_{a} (\tau, \sigma) +2\pi\, l^{2} \,  p_{a} \label{t33}
\end{eqnarray}
with $a= 0, \cdots, d$ and with $(w^{a}, lp_{a})$ being the components of a vector spanning a Lorentzian lattice $\Lambda^{d,d}$.

It is possible to reconduce the action (\ref{act1str}) to a sum of Floreanini-Jackiw Lagragians by block-diagonalizing simultaneosuly $C$ and $M$ through the matrix \cite{D}:
\begin{eqnarray}
({\cal T}^{-1})^{ij} = \frac{1}{\sqrt 2} \left(\begin{array}{cc}
                                               (G^{-1})^{\mu \nu} & (G^{-1})^{\mu \nu} \\
                                               (-E^{t} \,G^{-1})_{\mu}^{\,\, \nu} & (E\, G^{-1})_{\mu}^{\,\, \nu} \end{array}\right)~~~~  \label{calT}
\end{eqnarray}
where $E \equiv G + B$.
In fact, the matrix ${\cal T}^{-1}$ transforms $C$ and $M$  respectively into:
\begin{eqnarray}
{\cal T}^{-t} C {\cal T}^{-1}=  \left(\begin{array}{cc}
                       G^{-1}&0\\
                       0&-G^{-1}\end{array}\right)\equiv  {\cal C}^{-1} ~~; ~~{\cal T}^{-t}M {\cal T}^{-1}=  \left(\begin{array}{cc}
                       G^{-1}&0\\
                       0&G^{-1}\end{array}\right)\equiv {\cal G}^{-1}\label{trgo}
\end{eqnarray}
and introduces new {\em chiral coordinates} $\Phi_{i} = {\cal T}_{ij} \chi^{j} \equiv (X_{R\,\mu} , X_{L\, \mu})$,  in terms of which the $R$ and $L$ sectors  are completely decoupled also in the presence of the $B$-field. The matrix ${\cal G}^{-1} $ is the generalized metric in the chiral coordinates system.

In the flat gauge, previously introduced, the action becomes:
\begin{eqnarray}
S = \frac{T}{2}  \int d^{2} \xi \left[ \pm \frac{1}{2}  \partial_{0} X_{L; R}^{t} G^{-1} \partial_{1} X_{L; R}   - \frac{1}{2} \partial_{1} X_{L; R}^{t} G^{-1} \partial_{1} X_{L; R} \right] \label{FJrl}
 \label{actrl}
\end{eqnarray}
which is just the realization in the double string theory of the Floreanini-Jackiw Lagrangians (\ref{FJL}) with a nonvanishing Kalb-Ramond field as background.
Eqs. (\ref{condition}) and (\ref{fi0}) can be rewritten in terms of the Hodge duals of $dX_{R}$ and $dX_{L}$\footnote{The conventions used here for $p$-forms in a $D$-dimensional space-time with metric $G$ having signature $(-,+^{(D-1)})$ are the following: $w_{(n)}=\frac{1}{n} w_{\mu_{1} \dots \mu_{n}} dx^{\mu_{1}} \wedge dx^{\mu_{n}}$ and
$*w_{(n)}=\frac{\sqrt{-\mbox{det}G}}{n! (D-n)!} \epsilon_{\nu_{1} \dots \nu_{D-n} \mu_{1} \dots \mu_{n}} w^{\mu_{1} \dots \mu_{n}} dx^{\nu_{1}} \wedge dx^{\nu_{n}} $ with $\epsilon^{0\dots D-1}=1$.} as, respectively, the chirality and anti-chirality conditions: 
\begin{eqnarray}
*dX_R=dX_R ~~;~~*dX_L=-dX_L \,\, . \label{cfet}
\end{eqnarray}
The solution of eqs. (\ref{cfet}), with identifications on the torus now rewritten as:
\begin{eqnarray}
X_{R \, \mu}[\tau - (\sigma +\pi)] = X_{R \, \mu}(\tau - \sigma) -2\pi\, l^{2}\,  p_{R \, \mu} \label{cr} \\
X_{L \, \mu}[\tau + (\sigma +\pi)]= X_{L \, \mu}(\tau + \sigma) +2\pi\, l^{2}\,  p_{L \, \mu} \label{cl}
\end{eqnarray}
with
\begin{eqnarray}
\left(\begin{array}{c} -l p_R\\ l p_L\end{array}\right)={\cal T} \left(\begin{array}{c}w\\ lp\end{array}\right)\,\,,\label{nl}
\end{eqnarray}
 is given by:
\begin{eqnarray}
X_R (\tau - \sigma)&=& x_R+2\,l^2\,p_R(\tau -\sigma)+i l\sum_{n\neq 0}\frac{{\alpha}_n}{n}e^{-2i n(\tau- \sigma)} \label{xr_exp} \\
 X_L (\tau + \sigma) &=& x_L+2\,l^2\,p_L(\tau +\sigma)+i l\sum_{n\neq 0}\frac{\tilde{\alpha}_n}{n}e^{-2i n(\tau+\sigma) } \label{xl_exp}
\end{eqnarray}
formally identical to the usual expansion of the right and left bosonic string coordinates.

It is convenient to introduce, at this  point, the world-sheet light-cone coordinates $\sigma^+=\tau+\sigma$ and $\sigma^-=\tau-\sigma$. In terms of them, the components of the tensor $t^{\,\,\,a}_{b}$ turn out to be:
\begin{eqnarray}
t_{++} &=&  \partial_+X_R^tG^{-1} \partial_+ X_R + \partial_+X_L^t G^{-1} \partial_+X_L
- 2 \partial_+X_L^tG^{-1} \partial_-X_L \nonumber \\
\label{12cos}
t_{--} &=&  \partial_-X_R^tG^{-1} \partial_- X_R - \partial_+ X_L^t G^{-1} \partial_- X_L
- 2 \partial_+X_R^tG^{-1} \partial_-X_R   \label{prr}
\end{eqnarray}
while the Weyl invariance imposes $t_{+-} = -t_{-+}$, with
\begin{eqnarray}
t_{+-} = - \frac{1}{4} \epsilon^{ab} t_{ab} =\partial_- X_L^t G^{-1} \partial_- X_L - \partial_+ X_R^t G^{-1} \partial_+X_R   \label{3cos}
\end{eqnarray}
and   $\partial_{\pm} = \frac{1}{2} (\partial_{0} \pm \partial_{1})$.
The quantity defined in (\ref{3cos}) of course vanish on-shell, while the other two quantites in (\ref{prr}) have to be seen as contraints to be imposed at the classical and quantum level. 
In the following, the constraints are going to be evaluated on the solution of the equation of motion for $X_{R}$ and $X_{L}$, according the standard procedure followed in string theory.
These constraints, on-shell, become:
\begin{eqnarray}
 t_{++} &=&  \partial_+X_L^t G^{-1}\partial_+X_L\approx 0 \nonumber\\
t_{--} &=&  \partial_-X_R^t\,G^{-1}\partial_-X_R \approx 0  \nonumber.
\end{eqnarray}
and they look like the contraints for the energy-momentum tensor in the usual bosonic string theory leading to the Virasoro algebra. The constraint $t_{+-} \approx 0$ is already satisfied on-shell.
 
Having put our Lagrangian in the Floreanini-Jackiw form, it results to be linear in time derivatives, hence it has also primary constraints, besides the above mentioned constraints: \begin{eqnarray}
\Psi_R(P_R,\, X_R)=P_R + \frac{T}{2} G^{-1} \partial_1X_R\approx 0 ~~;~~ \Psi_L(P_L,\,X_L)=P_L- \frac{T}{2} G^{-1} \partial_1X_L\approx 0 \,\, . \label{clcos}
\end{eqnarray}
They are provided by the definition of the conjugate momenta:
\begin{eqnarray}
P_R \equiv \frac{\partial {\cal L}_R}{\partial (\partial_0 X_R^t)} = - \frac{T}{2} G^{-1} \partial_1 X_R ~~;~~ P_L \equiv \frac{\partial {\cal L}_L}{\partial (\partial_0 X_L^t)} = \frac{T}{2} G^{-1}  \partial_1 X_L . \label{defm}
\end{eqnarray}
The classical dynamic of the  system is studied by defining the Poisson brackets
\begin{eqnarray}
\left\{P_{R; L}(\tau,\,\sigma),\,X_{R; L}^{t}(\tau,\,\sigma')\right\}_{PB}= \delta(\sigma-\sigma') \mathbb{I} \label{pb}  \,\, .
\end{eqnarray}
According to the previous definition, the primary constraints satisfy the following algebra
\begin{eqnarray}
\left\{ \Psi_{R; L}(\tau,\sigma),\,\Psi_{R; L}^{t}(\tau, \sigma')\right\}_{PB} = \mp TG^{-1}\delta'(\sigma-\sigma')~~ \label{eq51}
\end{eqnarray}
with $\delta'(x)=\partial_x\delta(x)$ and the upper [lower] sign on the right hand side of the previous identity refers to the label $R$ [$L$] on the left  of the same  equation.  The algebra in eq. (\ref{eq51}) implies that these primary constraints are second class and therefore the Dirac quantization method has to be applied yielding:
\begin{eqnarray}
\left\{X_{R; L}(\tau,\,\sigma),\,X^{t}_{R; L}(\tau,\,\sigma')\right\}_{DB}&=&\mp \frac{G}{T}\epsilon(\sigma-\sigma')   \nonumber  \\
\left\{P_{R; L}(\tau,\,\sigma),\,X^{t}_{R; L}(\tau,\,\sigma')\right\}_{DB}&=&\frac{1}{2}\delta(\sigma-\sigma') {\mathbb I}   \label{RLDB}\\
\left\{P_{R; L}(\tau,\,\sigma),\,P^{t}_{R; L}(\tau,\,\sigma')\right\}_{DB}&=&\pm\frac{T}{4}G^{-1}\delta'(\sigma-\sigma') \nonumber
\end{eqnarray}
where $\epsilon(\sigma-\sigma')$ is the unit step function.

On the equation of motion, the algebra of the constraints reads as:
\begin{eqnarray}
\left\{ \Psi_{R} (\tau, \sigma), t_{--}(\tau, \sigma') \right\}_{PB}= \delta'(\sigma - \sigma') G^{-1} \partial_{-} X_{R}(\tau -\sigma) \approx 0
\end{eqnarray}
with a similar relation for $\Psi_{L}$ and $t_{++}$.  Here the last relation  comes from the constraint $t_{--} \approx 0$.
The first of the Dirac parenthesis in eq. (\ref{RLDB})  can be rewritten in terms of the  original variables $X$ and $\tilde{X}$ in  such a way to generalize the first commutator in eqs.  (\ref{comm}):
\begin{eqnarray}
&&\left\{X(\tau,\,\sigma),\,\tilde{X}^{t}(\tau,\,\sigma')\right\}_{DB}=\frac{1}{T}\epsilon(\sigma-\sigma') {\mathbb I}  \nonumber\\
&&\left\{{P}(\tau,\,\sigma),\,{X}^{t}(\tau,\,\sigma')\right\}_{DB}=\left\{\tilde{P}(\tau,\,\sigma),\,
\tilde{X}^{t}(\tau,\,\sigma')\right\}_{DB}=
\frac{1}{2}\delta(\sigma-\sigma') {\mathbb I} \label{RLDB1}
\\
&&\left\{{P}(\tau,\,\sigma),\,{\tilde P}^{t}(\tau,\,\sigma')\right\}_{DB}= -\frac{T}{4}\delta'(\sigma-\sigma') {\mathbb I} \nonumber
\end{eqnarray}
where $ P$ and $\tilde{P}$  are the conjugate momenta with respect to $X$ and $\tilde{X}$.
It is interesting to see that the commutators so obtained generate for the Fourier modes of the coordinates the standard commutators  in the usual string formulation \cite{DGMP}:

\begin{eqnarray}
[p_{R; L},\,x_{R; L}^t]= i G ~~;~~ [\alpha_m,\, {\alpha_n}^t]= m G\delta_{ n+m,0}~~;~~[\tilde{\alpha}_m,\,\tilde{\alpha}_n^t]= m G\delta_{ n+m,0}  \,\, . \label{fmcr}
\end{eqnarray}
These relations can be used in the constraint algebra and one recovers, in the R,L-sectors, the usual Virasoro algebra with vanishing quantum conformal anomaly in $D=26$. 

One can conclude that the non-covariant sigma model with an explicit $O(D,D)$ invariance is an extension of the usual formulation of the bosonic string theory obtained by {\em doubling}  the dimensions of the target space. The string coordinates on the doubled target space are non-commuting phase coordinates but {\em generate the usual commutators} for the Fourier modes.

\section{Covariant action (Hull): classical and quantum aspects}

 The  covariant manifestly T-dual invariant  formulation of the double string theory due to Hull \cite{H} is going to be discussed in this section. It is  
  defined by the coordinates $(Y(\tau,\,\sigma),\,\cal{X}(\tau,\,\sigma))$  mapping the string world-sheet in the target space. Locally, the target space looks like $\mathbb{R}^{1,  D-1}\otimes T^{2d}$ where the coordinates $Y \equiv (Y^{I})$ ($I=0, \dots, D-1$) are associated with the non-compact space-time  while the coordinates ${\cal X}\equiv ({\cal X}^i)$($i=1,\dots, 2d$), through the identification given in eq. (\ref{t33}), describe the double torus.

The world-sheet action proposed in ref. \cite{H} is
\begin{eqnarray}
S=- \frac{T}{4} \int 
d{\cal X}^i\,M_{ij}(Y)\wedge*d{\cal X}^j
\label{inva}
\end{eqnarray}
where $M$ is a generalized metric. 

The action, supplemented by the torus identifications given in eq. (\ref{t33}), is invariant under the   $GL(2d;\mathbb{Z})$ group which is the manifest symmetry of the theory \cite{H}. Since the number of the coordinates on the torus has been doubled {\em a priori}, a self-duality constraint that could halve them has to be imposed:
\begin{eqnarray}
 *M_{ij}\,d{\cal X}^j =  -\Omega_{ij}\, d{\cal X}^j\label{Hcos}\;\; .
 \label{gugo}
\end{eqnarray}
Here $\Omega$ is the $O(d,d)$ invariant metric defined in eq. (\ref{omega}). With this choice, the invariance of the theory reduces to the one under  $O(d,d; \mathbb{Z})$.  Eq. (\ref{gugo}) is identical to the constraint on the $\epsilon$-trace in the non-covariant formulation necessary for restoring, in that case, the Lorentz local invariance. More precisely, it reproduces the constraint:
\begin{eqnarray}
- \epsilon_{ab} C_{ij} \partial^{b} \chi^{j} + M_{ij} \partial_{a} \chi^{j} =0
\end{eqnarray}
which contains, in a covariant form, both  the equation of motion for $\chi^{i}$ in eq. (\ref{fi0}) and the constraint in eq. (\ref{condition}). It is interesting to observe here that the condition in eq. (\ref{gugo}) coincides with the one obtained in eq. (25) of ref. \cite{D} by uniting equations of motion and Bianchi identities into a single equation, in a context in which duality rotations of scalars relate two different sigma models.

The energy-momentum tensor obtained from this action turns out to be:
\begin{eqnarray}
T_{\alpha\beta}=-\frac{4}{T } \frac{1}{\sqrt {- g}} \frac{\delta S}{\delta g^{\alpha\beta}}=
\partial_{\alpha} {\cal \chi}^t
\,M\,\partial_\beta {\cal \chi} -\frac{1}{2} g_{\alpha\beta}\partial_\gamma{\cal \chi}^t\,M\,
\partial^{\gamma}{\cal \chi}   .
\end{eqnarray}
It is traceless because  of the Weyl invariance.  This, together with the invariance under reparametrizations of the world-sheet, is used to gauge-fix   the two-dimensional metric so that $g_{\alpha \beta} =\eta_{\alpha\beta}$.
The equations of motion for ${\cal \chi}$, clearly satisfied on the constraint surface, are:
\begin{eqnarray}
d*(Md{\cal \chi})=0
\label{eqmot1}
\end{eqnarray}
with boundary conditions given by the surface integral:
\begin{eqnarray}
-\frac{T}{2} \left. \int d\tau \, \delta {\cal X}^{t} M \partial_{1} {\cal X} \right|^{\sigma = \pi}_{\sigma=0}
\end{eqnarray}
which vanishes, in particular,  if periodicity conditions, peculiar of closed strings, are imposed.
As already done in the non-covariant formulation,
it is convenient to introduce the right and left coordinates $\Phi_i=(X_{R \, \mu},\,X_{L\, \mu})$:
\begin{eqnarray}
\Phi_i= {\cal T}_{ij}{\cal \chi}^j~~;~~
{\cal T} = \frac{1}{\sqrt 2} \left(\begin{array}{cc}
                                               E^t& \mathbb{I}\\
                                               E&-\mathbb{I}\end{array}\right) \,\, .
\end{eqnarray}
The action (\ref{inva}), when rewritten in terms of these coordinates, becomes:
 \begin{eqnarray}
S= -\frac{T}{4} \int  d\Phi ^t\,{\cal G}^{-1}\wedge*d\Phi \,\, 
\end{eqnarray}
where the generalized metric is given now by:
\[
{\cal G}^{-1} = \left( \begin{array}{cc}  G^{-1} & 0 \\ 0 & G^{-1} \end{array} \right) \,\, .
\]
Furthermore, in the new coordinate system, the constraints become the chirality and anti-chirality conditions
\begin{eqnarray}
\frac{2}{T}\,{\Psi}_{R} \equiv dX_R- *dX_R=0~;~~ \frac{2}{T}{\Psi}_{L} \equiv dX_L+*dX_L=0  \label{Hncos}
\end{eqnarray}
that generalize in this case the self-dual and anti-self dual constraints satisfied by the usual string coordinates compactified on a torus. Eqs. (\ref{Hncos}) formally determine four conditions for the $X_{R;L}$ coordinates. However, only two of them are independent and they  can be written in the following form:
\begin{eqnarray}
\left( {\Psi}_{R;L} \right)_{0} = \pm\left( {\Psi}_{R;L} \right)_{1} \equiv T\partial_\pm X_{R;L}= G P_{R;L}\pm\frac{T}{2} \partial_{1} X_{R;L} =0   \label{beh}
\end{eqnarray}
where the  definition of the conjugate momentum has been used. These constraints coincide with the  ones in eq.  (\ref{clcos}) and so
 satisfy the algebra given in eq. (\ref{eq51}), {\em behaving} like second-class constraints.
 
 In the  world-sheet light-cone coordinates $\sigma^{\pm}=\tau\pm \sigma$, the non-vanishing components of the energy-momentum tensor in these coordinates are:
\begin{eqnarray}
T_{++}=\frac{1}{2}\left( T_{00}+T_{01}\right)= \partial_+\Phi^t{\cal G}^{-1} \partial_+\Phi~~;~~
T_{--}=\frac{1}{2}\left( T_{00}-T_{01}\right)= \partial_-\Phi^t{\cal G}^{-1}\partial_-\Phi \,\, ,
\end{eqnarray}
being, as usual, $\partial_\pm=\frac{1}{2}(\partial_0\pm\partial_1)$. It is also useful to express  the
components of the energy-momentum tensor in terms of the constraints in eq. (\ref{beh}) as:
\begin{eqnarray}
T_{++} & = & \frac{1}{T^2}\Psi^{t}_RG^{-1}\Psi_R+\partial_+X^{t}_LG^{-1}\partial_+X_L \nonumber \\
\\
T_{--} &= & \frac{1}{T^2} \Psi^{t}_LG^{-1}\Psi_L+\partial_-X^{t}_RG^{-1}\partial_-X_R . \nonumber
\end{eqnarray}
It is easy to check that the left and right sectors commute by definition, while
\begin{eqnarray}
\left\{T_{\pm \pm},\, \Psi_{R,L}\right\}_{PB} = \mp \frac{2}{T} \delta' (\sigma - \sigma') \Psi_{R,L} \approx 0 \,\, ,
\end{eqnarray}
where the ``weak"  identity to zero is meant on the surface of the constraints. Furthermore, the following identity holds:
\begin{eqnarray}
 \left\{
\partial_{\mp}X_{R;L}(\tau,\,\sigma),\,\Psi_{R;L}(\tau,\,\sigma')\right\}=0 \,  .
\end{eqnarray}
Since the constraints in eq. (\ref{beh}) behave like second class constraints, they are treated by the Dirac method of quantization. The Dirac brackets between the canonical coordinates are:
\begin{eqnarray}
\left\{ P_{R;L}(\tau,\,\sigma),\,X^{t}_{R;L}(\tau,\,\sigma')\right\}_{DB}&=&\frac{1}{2}\, {\mathbb I}\,  \delta(\sigma-\sigma')\nonumber\\
\left\{ X_{R;L}(\tau,\,\sigma),\,X^{t}_{R;L}(\tau,\,\sigma')\right\}_{DB}&=&\mp\frac{G}{T}  \epsilon(\sigma-\sigma') \label{DCB} \\
\left\{ P_{R;L}(\tau,\,\sigma),\,P^{t}_{R;L}(\tau,\,\sigma')\right\}_{DB}&=&\pm\frac{T}{4}G^{-1} \delta'(\sigma-\sigma') \nonumber \,\, .
\end{eqnarray}
Furthermore, the above ``second class'' constraints can be strongly imposed implying
$X_R\equiv X_R(\sigma^-)$ and $X_L\equiv X_L(\sigma^+)$. These identities, once solved with the closed string boundary conditions, lead to the the Fourier expansions  given in eqs. (\ref{xr_exp}, \ref{xl_exp}).

The expression of the energy-momentum tensor on the surface constraint simplifies becoming:
\begin{eqnarray}
T_{++}= \partial_+ X_L^t{ G}^{-1} \partial_+X_L~~;~~
T_{--}= \partial_-X^t_R{ G}^{-1} \partial_-X_R \,\, .
\end{eqnarray}
Eq. (\ref{DCB}) determines the following Dirac brackets for  the coordinates Fourier modes:
\begin{eqnarray}
\{p_{R;L},\,x^{t}_{R;L}\}_{DB}= G ~;~\{ \alpha_n,\,\alpha^{t}_m \}_{DB}= i n \,G\,\delta_{n+m,0}~;~\{ \tilde{\alpha}_n,\,\tilde{\alpha}^{t}_{m} \}_{DB}=i n \,G\,\delta_{n+m,0} \label{dirbra}
\end{eqnarray}
which again coincide with the Poisson brackets of the string modes in the bosonic string theory.

 The quantization of this theory is therefore exactly the same as the
one of the Tseytlin model.

\section{Conclusion}

Two T-duality symmetric  formulations of bosonic string theory have been presented. In absence of interaction, they are equivalent at classical and quantum level providing a generalization of the usual theory. A doubling of the coordinates is required while the quantization generates a non-commuting geometry. 

\vspace{.5cm}

{\bf Acknowledgments}. The author thanks his collaborators in ref. \cite{DGMP} for sharing with them the results contained in this work. Furthermore, he is  deeply grafetul to the organizers of the  ``Workshop on Quantum Fields and Strings" (Corfu,  September 14-21, 2014), in particular to Ioannis Bakas and George Zoupanos,  for their warm hospitality and the great and stimulating atmosphere they managed to create.

\end{document}